\overfullrule=0pt
\input harvmac
\input epsf.sty

\def\a{{\alpha}}

\def\l{{\lambda}}

\def\b{{\beta}}

\def\g{{\gamma}}

\def\d{{\delta}}
\def\e{{\epsilon}}

\def\half{{1\over 2}}
\def\p{{\partial}}

\def\t{{\theta}}

\def\bar{\overline}
\def\({\left(}
\def\){\right)}

\def\cF{{\cal F}}
\def\ang#1{\left\langle #1 \right\rangle}



\Title{\vbox{\hbox{AEI-2009-097}}} {\vbox{
    \centerline{\bf Simplifying the Tree-level Superstring Massless Five-point Amplitude }
 }
}

\bigskip\centerline{Carlos R. Mafra\foot{email: crmafra@aei.mpg.de}}

\bigskip

\centerline{\it }
\medskip
\centerline{\it Max-Planck-Institut f\"ur
Gravitationsphysik, Albert-Einstein-Institut}
\smallskip
\centerline{\it 14476 Golm, Germany}
\medskip

\vskip .3in

We use the pure spinor formalism to obtain the supersymmetric massless five-point
amplitude at tree-level in a streamlined fashion. We also prove the
equivalence of an OPE identity in string theory with
a subset of the Bern-Carrasco-Johansson five-point kinematic relations, and
demonstrate how the remaining BCJ identities follow from the different integration 
regions over the open string world-sheet, therefore providing a first principles
derivation of the (supersymmetric) BCJ identities.

\vskip .3in

\Date {September 2009}

\lref\superpoincare{
  N.~Berkovits,
  ``Super-Poincare covariant quantization of the superstring,''
  JHEP {\bf 0004}, 018 (2000)
  [arXiv:hep-th/0001035].
}

\lref\multiloop{
  N.~Berkovits,
  ``Multiloop amplitudes and vanishing theorems using the pure spinor
  formalism for the superstring,''
  JHEP {\bf 0409}, 047 (2004)
  [arXiv:hep-th/0406055].
}
\lref\nonMin{
  C.~R.~Mafra and C.~Stahn,
  ``The One-loop Open Superstring Massless Five-point Amplitude with the
  Non-Minimal Pure Spinor Formalism,''
  JHEP {\bf 0903}, 126 (2009)
  [arXiv:0902.1539 [hep-th]].
}
\lref\MafraAR{
  C.~R.~Mafra,
  ``Pure Spinor Superspace Identities for Massless Four-point Kinematic
  Factors,''
  JHEP {\bf 0804}, 093 (2008)
  [arXiv:0801.0580 [hep-th]].
}

\lref\MedinaD{
  R.~Medina, F.~T.~Brandt and F.~R.~Machado,
  ``The open superstring 5-point amplitude revisited,''
  JHEP {\bf 0207}, 071 (2002)
  [arXiv:hep-th/0208121].
}
\lref\MedinaC{
  L.~A.~Barreiro and R.~Medina,
  ``5-field terms in the open superstring effective action,''
  JHEP {\bf 0503}, 055 (2005)
  [arXiv:hep-th/0503182].
}

\lref\medinafermions{
  R.~Medina and L.~A.~Barreiro,
  ``Higher N-point amplitudes in open superstring theory,''
  PoS {\bf IC2006}, 038 (2006)
  [arXiv:hep-th/0611349].
}

\lref\StiebergerBH{
  S.~Stieberger and T.~R.~Taylor,
  ``Amplitude for N-gluon superstring scattering,''
  Phys.\ Rev.\ Lett.\  {\bf 97}, 211601 (2006)
  [arXiv:hep-th/0607184].
}
\lref\stie{
  D.~Oprisa and S.~Stieberger,
  ``Six gluon open superstring disk amplitude, multiple hypergeometric  series
  and Euler-Zagier sums,''
  arXiv:hep-th/0509042.
}
\lref\stieMulti{
  S.~Stieberger and T.~R.~Taylor,
  ``Multi-gluon scattering in open superstring theory,''
  Phys.\ Rev.\  D {\bf 74}, 126007 (2006)
  [arXiv:hep-th/0609175].
}  
\lref\stieMHV{
  S.~Stieberger and T.~R.~Taylor,
  ``Supersymmetry Relations and MHV Amplitudes in Superstring Theory,''
  Nucl.\ Phys.\  B {\bf 793}, 83 (2008)
  [arXiv:0708.0574 [hep-th]].
}

\lref\vanhoveN{
  N.~E.~J.~Bjerrum-Bohr, P.~H.~Damgaard and P.~Vanhove,
  ``Minimal Basis for Gauge Theory Amplitudes,''
  arXiv:0907.1425 [hep-th].
}
\lref\bernN{
  Z.~Bern, J.~J.~M.~Carrasco and H.~Johansson,
  ``New Relations for Gauge-Theory Amplitudes,''
  Phys.\ Rev.\  D {\bf 78}, 085011 (2008)
  [arXiv:0805.3993 [hep-ph]].
}
\lref\PSS{
C.~R.~Mafra, ``PSS: A FORM program to compute pure spinor superspace expressions'',
work in progress.
}

\lref\vonkMHV{
  R.~Boels, K.~J.~Larsen, N.~A.~Obers and M.~Vonk,
  ``MHV, CSW and BCFW: field theory structures in string theory amplitudes,''
  JHEP {\bf 0811}, 015 (2008)
  [arXiv:0808.2598 [hep-th]].
}

\lref\anomalia{
  N.~Berkovits and C.~R.~Mafra,
  ``Some superstring amplitude computations with the non-minimal pure spinor
  formalism,''
  JHEP {\bf 0611}, 079 (2006)
  [arXiv:hep-th/0607187].
}

\lref\twoloop{
  N.~Berkovits,
  ``Super-Poincare covariant two-loop superstring amplitudes,''
  JHEP {\bf 0601}, 005 (2006)
  [arXiv:hep-th/0503197].
}

\lref\uloop{
  C.~R.~Mafra,
  ``Four-point one-loop amplitude computation in the pure spinor formalism,''
  JHEP {\bf 0601}, 075 (2006)
  [arXiv:hep-th/0512052].
}

\lref\brenno{
  N.~Berkovits and B.~C.~Vallilo,
  ``Consistency of super-Poincare covariant superstring tree amplitudes,''
  JHEP {\bf 0007}, 015 (2000)
  [arXiv:hep-th/0004171].
}

\lref\twolooptwo{
  N.~Berkovits and C.~R.~Mafra,
  ``Equivalence of two-loop superstring amplitudes in the pure spinor and  RNS
  formalisms,''
  Phys.\ Rev.\ Lett.\  {\bf 96}, 011602 (2006)
  [arXiv:hep-th/0509234].
}

\lref\stahn{
  C.~Stahn,
  ``Fermionic superstring loop amplitudes in the pure spinor formalism,''
  JHEP {\bf 0705}, 034 (2007)
  [arXiv:0704.0015 [hep-th]].
}

\lref\nekrasov{
  N.~Berkovits and N.~Nekrasov,
  ``Multiloop superstring amplitudes from non-minimal pure spinor formalism,''
  JHEP {\bf 0612}, 029 (2006)
  [arXiv:hep-th/0609012].
}
\lref\yuri{
  Y.~Aisaka and N.~Berkovits,
  ``Pure Spinor Vertex Operators in Siegel Gauge and Loop Amplitude
  Regularization,''
  JHEP {\bf 0907}, 062 (2009)
  [arXiv:0903.3443 [hep-th]].
}
\lref\grassi{
  P.~A.~Grassi and P.~Vanhove,
  ``Higher-loop amplitudes in the non-minimal pure spinor formalism,''
  JHEP {\bf 0905}, 089 (2009)
  [arXiv:0903.3903 [hep-th]].
}
\lref\theorems{
  N.~Berkovits,
  ``New higher-derivative R**4 theorems,''
  Phys.\ Rev.\ Lett.\  {\bf 98}, 211601 (2007)
  [arXiv:hep-th/0609006].
}
\lref\green{
  N.~Berkovits, M.~B.~Green, J.~G.~Russo and P.~Vanhove,
  ``Non-renormalization conditions for four-gluon scattering in supersymmetric
  string and field theory,''
  arXiv:0908.1923 [hep-th].
}
\lref\purified{
  G.~Policastro and D.~Tsimpis,
  ``R**4, purified,''
  Class.\ Quant.\ Grav.\  {\bf 23}, 4753 (2006)
  [arXiv:hep-th/0603165].
}
\lref\nineloop{
  M.~B.~Green, J.~G.~Russo and P.~Vanhove,
  ``Ultraviolet properties of maximal supergravity,''
  Phys.\ Rev.\ Lett.\  {\bf 98}, 131602 (2007)
  [arXiv:hep-th/0611273].
}

\lref\PSSuperspace{
  N.~Berkovits,
  ``Explaining pure spinor superspace,''
  arXiv:hep-th/0612021.
}

\lref\tese{
  C.~R.~Mafra,
  ``Superstring Scattering Amplitudes with the Pure Spinor Formalism,''
  arXiv:0902.1552 [hep-th].
}

\lref\stieb{
  S.~Stieberger,
  ``Open \& Closed vs. Pure Open String Disk Amplitudes,''
  arXiv:0907.2211 [hep-th].
}
\lref\FORM{
  J.~A.~M.~Vermaseren,
  ``New features of FORM,''
  arXiv:math-ph/0010025.
\semi
  M.~Tentyukov and J.~A.~M.~Vermaseren,
  ``The multithreaded version of FORM,''
  arXiv:hep-ph/0702279.
}

\newsec{Introduction}

Since the discovery of the pure spinor formalism \superpoincare\ a new
method to efficiently compute supersymmetric scattering amplitudes 
is available.
Although its simplifying features manifest themselves more vividly 
in explicit one- and two-loop computations \refs{\twoloop,\twolooptwo,\multiloop,\uloop,\MafraAR} and
provide hope\foot{Useful knowledge can be obtained 
even without fully explicit higher-loop computations
\refs{\theorems,\green,\nineloop}.} for higher-loop extensions 
\refs{\nekrasov,\yuri,\grassi}, tree-level amplitudes \brenno\ also benefit from 
the streamlined nature of the formalism\foot{For a
review of scattering amplitudes in the pure spinor formalism, see \tese.}. 
In particular, having results written in terms of pure spinor superspace 
expressions \refs{\PSSuperspace,\MafraAR}
sheds new light into finding supersymmetric completions \purified.

This paper simplifies the long (bosonic) RNS five-point computations of \refs{\MedinaD,\MedinaC}
while naturally extending them to the full supersymmetric multiplet using the pure 
spinor superspace. In doing that we uncover the superstring origin of
the Bern-Carrasco-Johansson (BCJ) kinematic identities of \bernN, proving that some
of them come from an OPE identity and that they are supersymmetric. And in view of the
string theory proof for the four-point BCJ identity we will demonstrate that the remaining
BCJ relations follow from the different integration regions of the open string
world-sheet. 
These integrations over the various domain of integrations are related to the monodromy
identities \refs{\vanhoveN,\stieb} between the string theory amplitudes which have been used in \vanhoveN\
to prove that the number of partial amplitudes is $(N-3)!$, and is ultimately related 
to the BCJ identities of \bernN.

In the following, the open string massless five-point amplitude at tree-level will be shown
to be
${\cal A}_5 = \sum_{\rm non-cyclic} \tr(\l^{a_{i_1}}{\ldots} \l^{a_{i_5}})A(i_1,{\ldots},i_5)$,
with
\eqn\main{
A(1,2,3,4,5) = 
{{\tilde L}_{2131}\over \a_{12}\a_{45}} S_1
- {{\tilde L}_{2334}\over \a_{23}\a_{51}} S_2
- {{\tilde L}_{2134}\over \a_{12}\a_{34}} S_3
- {{\tilde L}_{2331}\over \a_{23}\a_{45}} S_4
- {{\tilde L}_{3424}\over \a_{34}\a_{51}} S_5
+ L_{2431}\, K_3(2\a')^2
}
where
$$
S_1 = T - (2\a')^2K_3(\a_{34}\a_{45} + \a_{23}\a_{12}), \quad
S_2 = T - (2\a')^2 K_3(\a_{12}\a_{23} + \a_{51}(\a_{45} + \a_{23}))
$$
$$
S_3 =  T - (2\a')^2K_3(\a_{34}\a_{45} + \a_{12}(\a_{51} + \a_{34})), \quad
S_4 = T - (2\a')^2 K_3(\a_{12}\a_{23} + \a_{45}(\a_{51} + \a_{23}))
$$
$$
S_5 = T - (2\a')^2 K_3(\a_{12}\a_{51} + \a_{23}\a_{34}),
$$
and $T = 1 + O(k^6) + {\ldots} $ and $K_3 = \zeta(2) + O(k^2) + {\ldots} $
have well-known expansions in terms of $\a_{ij} = (k^i\cdot k^j)$ \MedinaC. 
In the field theory limit we set $\a'\rightarrow 0$ and therefore $S_j \rightarrow 1$.  The
kinematic factors ${\tilde L}_{ijkl}$ are given by simple pure spinor superspace expressions
which satisfy the supersymmetric BCJ relations,
\eqn\explibcj{
{\tilde L}_{2331} = L_{3121} - {\tilde L}_{2131}, \quad {\tilde L}_{2334} = {\tilde L}_{3424} - L_{2434}.
}

The paper is organized as follows. In section 2 we compute the five-point amplitude
at tree-level and express it in terms of simple pure spinor superspace expressions.
In section 3 we prove an OPE identity from which the supersymmetric generalizations
for some of the BCJ relations can be obtained. Furthermore, using the analogy with the four-point amplitude 
derivation of the BCJ identity
we show how to obtain the remaining ones. The pure spinor
superspace computations are presented in Appendix A, together with the explicit 
proof of \explibcj\ directly in superspace. Of particular importance is the simplified
expression for the OPE of two integrated vertices presented in (A.2).
The Appendix B is devoted to writing down
an ansatz for a simplified expression of $A_{F^4}(\t)$, whose {\it bosonic}
component expansion agrees with the expression obtained in section 2.
The Appendix C is a formal rewriting of the ten-dimensional results using the
four-dimensional spinor helicity formalism, and we show agreement with the expressions
of \refs{\StiebergerBH,\stieMulti,\stieMHV, \vonkMHV}.
Finally, in Appendix D we derive the relations obeyed by the integrals $K_j$
which were used in section 2.

\newsec{The five point amplitude in pure spinor superspace}

\noindent Following the tree-level prescription of \superpoincare\ the open superstring 
5-point amplitude is
$$
{\cal A}_5 = \sum_{\rm non-cyclic} \tr(\l^{a_{i_1}}{\ldots} \l^{a_{i_5}})A(i_1,{\ldots},i_5)
$$
where the partial amplitude $A_5(1,2,3,4,5)$ is given by
\eqn\ampli{
A_5(1,2,3,4,5) = \langle (\l A^1)(z_1)(\l A^4)(z_4)(\l A^5)(z_5)\int dz_2 U^2(z_2)\int dz_3 U^3(z_3)\rangle.
}
The $SL(2,R)$ symmetry of the disc requires the fixing of three positions, chosen as
$(z_1, z_4, z_5) =(0,1, \infty)$. Therefore the integrals are over the region $0\le z_2\le z_3\le 1$.

Using the OPEs of the pure spinor formalism to integrate out the conformal weight-one variables,
\ampli\ assumes the following form
\eqn\expre{
A_5 = \int dz_2 dz_3 \prod_{i < j}|z_{ij}|^{2 \a' k_i\cdot k_j}\Big[ {L_{2131}\over z_{21}z_{31}}
+ {L_{2134}\over z_{21}z_{34}} - {L_{2434}\over z_{24}z_{34}} - {L_{2431}\over z_{24}z_{31}}
+ {L_{2331}\over z_{23}z_{31}} - {L_{2334}\over z_{23}z_{34}} 
+ {L_{2314}\over z_{23}^2}\Big]
}
where the kinematic factors $L_{ijkl}$ are given by the following pure
spinor superspace expressions (from now on we set $2\alpha' = 1$),
$$       
L_{2134} =
 \langle \big[ A^1_m(\l\g^{m}W^2) + (\l A^1)(k^1\cdot A^2)\big]\big[ A^4_n(\l\g^{n}W^3) + (\l A^4)(k^4\cdot A^3)\big](\l A^5)\rangle
$$
$$ +  \langle (k^1\cdot k^2)(A^4W^3)(\l A^1)(\l A^2)(\l A^5)\rangle
   + \langle (k^3\cdot k^4)(A^1W^2)(\l A^3)(\l A^4)(\l A^5) \rangle
$$
$$
L_{2131} = 
+ \langle \big[ A^1_m (\l\g^{m}W^2) + (\l A^1)(k^1\cdot A^2)\big](\l A^4)(\l A^5)((k^1+k^2)\cdot A^3)\rangle
$$
$$ 
+ \langle \big[  A^1_m(\l\g^{m}W^3)(k^1\cdot A^2) -  A^{1\,m}(\l\g^{n}W^3){\cal F}^2_{mn}
       -  (\l \g^{m}W^3)(W^1\g^{m}W^2)\big](\l A^4)(\l A^5) \rangle
$$
$$ 
+ \langle \big[ (k^1\cdot k^2)(A^1W^3)(\l A^2)
       -  (k^1\cdot k^2)(A^2W^3)(\l A^1)\big](\l A^4)(\l A^5) \rangle
$$
$$
       + \langle  \big[ (k^1\cdot k^3)(A^1W^2)(\l A^3)
       +  (k^2\cdot k^3)(A^1W^2)(\l A^3)\big] (\l A^4)(\l A^5) \rangle
$$
$$
L_{2331} = 
 \langle  \big[  A^1_m(\l\g^{m}W^3) (k^3\cdot A^2)
         + { 1\over 4}(\l\g^{p}\g^{m n}W^3)A^1_p \cF^2_{m n}
    \big] (\l A^4)(\l A^5) \rangle
$$
$$  + \langle \big[ (\l A^1)(k^1\cdot A^3) (k^3\cdot A^2)
    + \half k^1_m (\l A^1)(W^2\g^{m}W^3) \big] (\l A^4)(\l A^5)\rangle
$$
\eqn\Ls{
+ \langle \big[ (k^2\cdot k^3)(A^1W^3)(\l A^2) 
         + (k^1\cdot k^2)(A^2W^3)(\l A^1) \big](\l A^4)(\l A^5) \rangle
     - (2\leftrightarrow 3)
}
while the other $L_{ijkl}$ are obtained by exchanging labels appropriately.
All the terms containing factors of
$(k^i\cdot k^j)(A^k W^l)$ are ``total derivative'' terms and will be 
shown to cancel in the final result.
Furthermore, the double pole 
in the OPE of $U^2(z_2)U^3(z_3)$ gives rise to the following expression for $L_{2314}$
\eqn\nontr{
L_{2314} =  (\a_{23} + 1)\langle \big[(A^2W^3) + (A^3W^2) - (A^2\cdot A^3)\big](\l A^1)(\l A^4)(\l A^5)\rangle
\equiv (\a_{23} + 1)L_{23}.
}
As will become clear later, the factor of $(1+\a_{23})$ appearing in \nontr\
is essential to obtain a simple answer for the amplitude. That this is possible
can be traced back to the fact that
the pure spinor Lorentz currents have level $-3$ (see the computations of Appendix A).

\noindent With the notation of \MedinaD\ for the integrals appearing\foot{The
RNS computations of \MedinaD\MedinaC\ required the evaluation of more complicated integrals
with cubic terms in the denominators.}
in \expre, the amplitude can be written as
${\cal A}_5 =  L_{2131} K_1 - L_{2134}K_2 - L_{2434}K'_1 + L_{2431}K_3
- L_{2331}K_5 - L_{2334}K'_4 + L_{2314}K_6$, or
\eqn\manifest{
A_5 = L_{2431}K_3  - L_{2134}K_2 + L_{2131} K_4 - L_{3424}K'_4
+ L_{2434}K'_5 - L_{3121}K_5  + L_{2314}K_6,
}
where we used $K_1 = K_4 - K_5$, $K'_1 = K'_4 - K'_5$ \MedinaC\ and
\eqn\depois{
L_{2331} = L_{3121} - L_{2131}, \quad L_{2334} = L_{3424} - L_{2434},
}
where \depois\ will be proved as an OPE identity in the next section.
Plugging in the expressions for $K_j$ in terms of $T$ and $K_3$ derived in
the Appendix D, the amplitude \manifest\
becomes
\eqn\amp{
{\cal A}_5(1,2,3,4,5) = T\, A_{\rm YM}(\t) + K_3\, A_{F^4}(\t),
}
where $A_{YM}(\t)$ and $A_{F^4}(\t)$ are superfields, 
\eqn\AYM{
A_{\rm YM}(\t) = {L_{2131}\over \a_{12}\a_{45}} - {L_{3424}\over \a_{34}\a_{51}} 
- {L_{2334}\over \a_{23}\a_{51}} - {L_{2331}\over \a_{23}\a_{45}}
- {L_{2134}\over \a_{12}\a_{34}}
- {L_{23}\over \a_{23}} \( {\a_{24}\over \a_{51}} + {\a_{13}\over \a_{45}} - 1\)
}
$$
A_{F^4}(\t) = 
L_{2431} - L_{2331} - L_{2334} - L_{2134}
+ {L_{23} \over \a_{23}}\(
 \a_{13}\a_{24}
- \a_{12}\a_{34}
- \a_{23}\a_{34}
- \a_{12}\a_{23}
\)
$$
$$
+ L_{2331}\({\a_{12}\over \a_{45}} + {\a_{51}\over \a_{23}} \)
+ L_{2334}\({\a_{34}\over \a_{51}} + {\a_{45}\over \a_{23}}  \)
+ L_{2134}\({\a_{45}\over \a_{12}} + {\a_{51}\over \a_{34}} \)
$$
\eqn\AFq{
+ L_{3424}\({\a_{12}\over \a_{34}} + {\a_{23}\over \a_{51}} \)
- L_{2131}\({\a_{34}\over \a_{12}} + {\a_{23}\over \a_{45}} \)
+ L_{23}\( {\a_{12}\a_{13}\over \a_{45}} + {\a_{34}\a_{24}\over \a_{51}}\).
}
From \AYM\ and \AFq\ all the other partial amplitudes can be obtained by permutation.
Therefore \amp\ is the supersymmetric generalization of equation (4.13) of \MedinaC.
In the field theory limit the amplitude \amp\ reduces to $A_{\rm YM}(\t)$.

Using the superspace expressions \Ls\ and \nontr\ one sees that
all terms containing factors of $(k^i\cdot k^j)(A^k W^l)$ cancel out in \AYM\ and \AFq.
For example, the terms in \AFq\ containing $(A^3 W^2)$ are given by
\eqn\totder{
\( {\a_{34}\over \a_{51}}(\a_{23}+\a_{24}+\a_{34}) - \a_{34}\)
\langle (A^3 W^2)(\l A^1)(\l A^4)(\l A^5)\rangle = 0
}
because $\a_{23}+\a_{24}+\a_{34} = \a_{51}$. In fact they come from total
derivative terms, as can be seen in the explicit computations of Appendix A.

The expressions \AYM\ and \AFq\ can be further simplified by 
absorbing the ``contact terms'' containing $L_{23}$ conveniently, taking
the relations \depois\ as a guide. Using the
identities $L_{3224} = - L_{2334}$ and $L_{3221} = - L_{2331}$, which follow
trivially from the antisymmetry of the simple pole of $U^2(z_2)U^3(z_3)$ under
$2 \leftrightarrow 3$, one can
can rewrite $A_{\rm YM}(1,2,3,4,5)$ and $A_{\rm YM}(1,3,2,4,5)$ as
\eqn\AYMd{
A_{\rm YM}(1,2,3,4,5) = {{\tilde L}_{2131}\over \a_{12}\a_{45}} - {{\tilde L}_{3424}\over \a_{34}\a_{51}} 
- {{\tilde L}_{2334}\over \a_{23}\a_{51}} - {{\tilde L}_{2331}\over \a_{23}\a_{45}}
- {{\tilde L}_{2134}\over \a_{12}\a_{34}}
}
\eqn\AYMt{
A_{\rm YM}(1,3,2,4,5) = {L_{3121}\over \a_{13}\a_{45}} - {L_{2434}\over \a_{24}\a_{51}} 
+ {{\tilde L}_{2334}\over \a_{23}\a_{51}} + {{\tilde L}_{2331}\over \a_{23}\a_{45}}
- {L_{3124}\over \a_{13}\a_{24}}
}
where we used $\a_{24} = \a_{51} - \a_{23} - \a_{34}$ and $\a_{13} = \a_{45} - \a_{23} - \a_{12}$ and
the redefined ${\tilde L}_{ijkl}$ are given by
$$
{\tilde L}_{2131} = L_{2131} + \a_{12}L_{23} - \a_{45}L_{23} ,\quad {\tilde L}_{3424} = L_{3424} - \a_{34}L_{23},
\quad {\tilde L}_{2134} = L_{2134} - \a_{34}L_{23}
$$
\eqn\redef{
{\tilde L}_{2334} = L_{2334} - \a_{34}L_{23},\quad {\tilde L}_{2331} = L_{2331} - \a_{12}L_{23} + \a_{45}L_{23}.
}
The identities \depois\ continue to hold with these redefinitions, that is
\eqn\BCJPS{
{\tilde L}_{2331} = L_{3121} - {\tilde L}_{2131}, \quad {\tilde L}_{2334} = {\tilde L}_{3424} - L_{2434}.
}
The use of \redef\ also removes the contact terms appearing in \AFq, simplifying it.
In fact, using \redef\ the supersymmetric string
theory partial amplitude \ampli\ becomes,
$$
{\cal A}_5(1,2,3,4,5) =
- {{\tilde L}_{2134}\over \a_{12}\a_{34}}\( T - K_3(\a_{34}\a_{45} + \a_{12}(\a_{51} + \a_{34}))\)
$$
$$
+ {{\tilde L}_{2131}\over \a_{12}\a_{45}}\( T - K_3(\a_{34}\a_{45} + \a_{23}\a_{12})\)
- {{\tilde L}_{3424}\over \a_{34}\a_{51}}\( T - K_3(\a_{12}\a_{51} + \a_{23}\a_{34})\)
+ L_{2431}\, K_3
$$
$$
- {{\tilde L}_{2331}\over \a_{23}\a_{45}}\( T - K_3(\a_{12}\a_{23} + \a_{45}(\a_{51} + \a_{23}))\)
- {{\tilde L}_{2334}\over \a_{23}\a_{51}}\( T - K_3(\a_{12}\a_{23} + \a_{51}(\a_{45} + \a_{23})\)
$$

The component expansions of \AFq\ and \AYM\ can be computed\foot{This task can
be implemented in a computer program \PSS\FORM.} using the methods of
\refs{\anomalia,\twolooptwo,\stahn}. When all external states are bosonic the
RNS results of \refs{\MedinaD,\MedinaC} are recovered,
\eqn\equivalente{
A_{\rm YM}(\t)\Big|_{NS} = - {1\over 2880} A^{\rm RNS}_{\rm YM}, 
\quad A_{F^4}(\t)\Big|_{NS} = - {1\over 2880} A^{\rm RNS}_{F^4}.
}
The higher $\alpha'$ expansion in \amp\ is determined solely by 
the expansions of $T$ and
$K_3$, and all the (supersymmetric) information about the external states is
encoded in the superfield expressions $A_{\rm YM}(\t)$ and $A_{F^4}(\t)$,
in accord with the observations of \medinafermions.
This is in fact a generic feature of the amplitudes computed in the 
pure spinor formalism. The kinematic factors of bosonic and fermionic
states are always multiplied by the same ``form factors'', which
come from the integrals over the world-sheet.

\newsec{Derivation of the BCJ kinematic identities}

In reference \bernN, the massless four-point partial amplitudes at
tree-level were represented in terms of its poles as
\eqn\quatrobcj{
A(1,2,3,4) = {n_s\over s} + {n_t\over t}, \quad A(1,3,4,2) = -{n_u\over u} - {n_s\over s},\quad
A(1,4,2,3) = -{n_t\over t} + {n_u\over u},
}
and the identity $n_u = n_s - n_t$ was explicitly shown to be true. Furthermore,
the five-point amplitudes were written as
$$
A_{\rm YM}(1,2,3,4,5) = {n_1\over \a_{12}\a_{45}} + {n_2\over \a_{23}\a_{51}} 
+ {n_3\over \a_{34}\a_{12}}
+ {n_4\over \a_{23}\a_{45}}
+ {n_5\over \a_{51}\a_{34}}
$$
$$
A_{\rm YM}(1,3,2,4,5) = {n_{15}\over \a_{13}\a_{45}} - {n_2\over \a_{23}\a_{51}} 
- {n_{10}\over \a_{24}\a_{13}}
- {n_4\over \a_{23}\a_{45}}
- {n_{11}\over \a_{51}\a_{24}}
$$
$$
A_{\rm YM}(1,4,3,2,5) = {n_{6}\over \a_{14}\a_{25}} + {n_5\over \a_{34}\a_{51}} 
+ {n_{7}\over \a_{23}\a_{14}}
+ {n_8\over \a_{25}\a_{34}}
+ {n_{2}\over \a_{51}\a_{32}}
$$
$$
A_{\rm YM}(1,3,4,2,5) = {n_{9}\over \a_{13}\a_{25}} - {n_5\over \a_{34}\a_{51}} 
+ {n_{10}\over \a_{24}\a_{13}}
- {n_8\over \a_{25}\a_{34}}
+ {n_{11}\over \a_{51}\a_{24}}
$$
$$
A_{\rm YM}(1,2,4,3,5) = {n_{12}\over \a_{12}\a_{35}} + {n_{11}\over \a_{24}\a_{51}} 
- {n_{3}\over \a_{34}\a_{12}}
+ {n_{13}\over \a_{35}\a_{24}}
- {n_{5}\over \a_{51}\a_{34}}
$$
\eqn\fiveptbcj{
A_{\rm YM}(1,4,2,3,5) = {n_{14}\over \a_{14}\a_{35}} - {n_{11}\over \a_{24}\a_{51}} 
- {n_{7}\over \a_{23}\a_{14}}
- {n_{13}\over \a_{35}\a_{24}}
- {n_{2}\over \a_{51}\a_{23}}
}
and by analogy with the Jacobi-like four-point kinematic relation, the numerators were required to
satisfy 
$$
n_3 - n_5 + n_8 = 0, \quad n_3 - n_1 + n_{12} = 0, 
\quad n_4 - n_1 + n_{15} = 0,\quad n_4 - n_2 + n_7 = 0,
$$
$$
n_5 - n_2 + n_{11} = 0, \quad n_7 - n_6 + n_{14} = 0, 
\quad n_8 - n_6 + n_{9} = 0,\quad n_{10} - n_9 + n_{15} = 0,
$$
\eqn\theids{
n_{10} - n_{11} + n_{13} = 0, \quad n_{13} - n_{12} + n_{14} = 0,
}
which was explicitly verified to be true.
Extending the same
reasoning to higher points, it was argued that those kind of relations
impose additional constraints which reduce the number of
independent $N$-point color-ordered amplitudes to $(N-3)!$. This conclusion
was later demonstrated in \vanhoveN\ using the field theory limit of string theory. 
We will now prove the identity \depois\ and discuss its relation\foot{I thank Pierre Vanhove for
several discussions about this.} with the 5-point BCJ identities
of \bernN.

To prove \depois\ it suffices to note that in the computation of
$$
\langle V^1(z_1)V^4(z_4)V^5(z_5)U^2(z_2)U^3(z_3)\rangle,
$$
a kinematic identity can be obtained by considering
the different orders in which the OPE's are evaluated. By
computing first the OPE's of $U^2(z_2)$ followed by $U^3(z_3)$ one gets,
\eqn\ones{
{L_{2131}\over z_{21}z_{31}} +
{L_{2134}\over z_{21}z_{34}} - {L_{2434}\over z_{24}z_{34}} 
- {L_{2431}\over z_{24}z_{31}}
+ {L_{2331}\over z_{23}z_{31}}
- {L_{2334}\over z_{23}z_{34}} + {L_{2314}\over z_{23}^2}
}
while in reverse order,
\eqn\twos{
{L_{3121}\over z_{31}z_{21}} 
+ {L_{3124}\over z_{31}z_{24}}
- {L_{3424}\over z_{34}z_{24}}
- {L_{3421}\over z_{34}z_{21}}
+ {L_{3221}\over z_{32}z_{21}}
- {L_{3224}\over z_{32}z_{24}} + {L_{3214}\over z_{32}^2}.
}
As the integrated vertex $U^I$ is bosonic, \ones\ and \twos\ must be equal. Therefore we get
$$
{(L_{2131} - L_{3121}) \over z_{21}z_{31}} - {(L_{2434} - L_{3424})\over z_{24}z_{34}} + 
L_{2331}({1 \over z_{23}z_{31}} + {1 \over z_{32}z_{21}})
- L_{2334}( {1\over z_{23}z_{34}} + {1\over z_{32}z_{24}})
$$
\eqn\used{
+ {1\over z_{34}z_{24}}(L_{2134} + L_{3421})
- {1\over z_{24}z_{31}}(L_{2431} + L_{3124}) = 0,
}
where we used $L_{3221} = - L_{2331}$ and $L_{2314} = L_{3214}$. To see this one notes that
$$
\langle [[U^2(z_2)U^3(z_3)] V^1(z_1)V^4(z_4)V^5(z_5)]\rangle = \langle [[U^3(z_3)U^2(z_2)] V^1(z_1)
V^4(z_4)V^5(z_5)]\rangle
$$
implies
$$
\lim_{z_2\rightarrow z_3}\big[ {L_{2331}\over z_{23}z_{31}} + {L_{2314}\over z_{23}^2}\big]
= \lim_{z_3\rightarrow z_2}\big[ {L_{3221}\over z_{32}z_{21}} + {L_{3214}\over z_{32}^2}\big]
$$
and therefore $L_{2331} = - L_{3221}$ and $L_{2314} = L_{3214}$. That is, the simple and double poles
of the $U^2(z_2)U^3(z_3)$ OPE are antisymmetric and symmetric under $ 2\leftrightarrow 3$,
respectively.
Finally, using
${1 \over z_{23}z_{31}} + {1 \over z_{32}z_{21}} = {1 \over z_{21}z_{31}}$ in \used\ leads to
\eqn\qed{
L_{2131} - L_{3121} + L_{2331} = 0, \quad L_{2434} - L_{3424} + L_{2334} = 0,
}
\eqn\qedd{
L_{2134} = - L_{3421}, \quad L_{2431} =- L_{3124}.
}
The identities \qed\ and \qedd\ can be also verified from their explicit
pure spinor superspace expressions given in the Appendix A.

After absorbing the contact terms as in \redef, the
field theory limit of the string partial amplitudes $A(1,2,3,4,5)$ and $A(1,3,2,4,5)$
are given by \AYMd\ and \AYMt, respectively. Note that there is an ambiguity (or freedom)
on how to absorb the contact terms, as there is no unique way in doing so. 
We chose to
absorb them while preserving the kinematic identities\foot{The full string
theory computation provides one extra layer of motivation for the redefinitions of \redef,
as they also remove the contact terms from the stringy correction $A_{F^4}(\t)$.}
\qed. This is in agreement
with the discussions of \bernN, where it is emphasized that the BCJ identities would not
be satisfied by any choice of absorbing contact terms. 

Comparing \AYMd\ and \AYMt\ with \fiveptbcj\ allow us to identify
$$
n_1 = {\tilde L}_{2131}, \quad n_4 = - {\tilde L}_{2331}, \quad n_{15} = L_{3121},\quad
n_5 = - {\tilde L}_{3424}, \quad n_2 = -{\tilde L}_{2334}, 
$$
\eqn\mapping{
\quad n_{11}  = L_{2434} = - L_{2443}, \quad
n_3 = - {\tilde L}_{2134}, \quad n_{10} = L_{3124} = - L_{2431}.
}
where $L_{2443} = - L_{2434}$  follows from
$ \langle [[U^2(z_2)V^4(z_4)] U^3(z_3)] \rangle = \langle [U^3(z_3) [U^2(z_2)V^4(z_4)] \rangle$.
Therefore \BCJPS\ is the supersymmetric generalization of 
the BCJ relations 
\eqn\psbcj{
n_4 - n_1 + n_{15} = 0, \quad n_5 - n_2 + n_{11} = 0.
}
Using $U^2$ and $U^4$ (or $U^3$ and $U^4$) as integrated vertices whose
positions run between 0 and 1 would lead to the BCJ identities
$n_{14} + n_{13} - n_{12} = 0$ and $n_5 - n_2 + n_{11} = 0$
(or $n_8 - n_6 + n_9 = 0$ and $n_5 - n_2 + n_{11} = 0$).

How can the remaining (supersymmetric) BCJ relations be obtained?
The four-point derivation of the BCJ identity provides the hint,
as there are no two integrated vertices to allow an OPE identity
in this case\foot{I thank Pierre Vanhove for discussions
on this point. In particular I acknowledge the fact that he kindly 
shared some notes where he suggested the relevance of the different
regions of integrations for obtaining the remaining BCJ ids. {\it Merci beaucoup, Pierre}!}.
Using the results of \MafraAR\
and the gamma function identity of $\Gamma(1+x) = x\Gamma(x)$ one can obtain
the open string partial amplitudes from $\langle V^1(0)\int U^2 V^3(1) V^4(\infty)\rangle$
by explicitly computing the integral over the three domains $0 \le z_2 \le 1$, $-\infty \le z_2 \le 0$,
and $1 \le z_2 \le \infty$,
$$
A(1,2,3,4) = \( - { K_{21} \over s} + {K_{23}\over t}\) {\Gamma(1-t)\Gamma(1-s)\over \Gamma(1+u)}
$$
$$
A(1,3,4,2) = \( - { tK_{21} \over su} + {K_{23}\over u}\) {\Gamma(1-s)\Gamma(1-u)\over \Gamma(1+t)}
$$
\eqn\FptBCJ{
A(1,3,2,4) = \( - { K_{21} \over u} + {s K_{23}\over ut}\) {\Gamma(1-t)\Gamma(1-u)\over \Gamma(1+s)}
}
where
$$
K_{21} = - \langle \big[ A^1_m (\l\g^m W^2) + (\l A^1)(k^1\cdot A^2)\big] (\l A^3)(\l A^4)\rangle
$$
$$
K_{23} = + \langle\big[ A^3_m (\l\g^m W^2) + (\l A^3)(k^3\cdot A^2)\big] (\l A^1)(\l A^4)\rangle
$$
and $s = - 2(k^1\cdot k^2) = -2 (k^3\cdot k^4)$, $u = - 2(k^1\cdot k^3) = - 2(k^2\cdot k^4)$,
$t = - 2(k^1\cdot k^4) = - 2(k^2\cdot k^3)$. Using  $s+t+u=0$ and taking the field theory limit
one can easily derive
the supersymmetric generalization of the four-point BCJ relation 
$n_u = n_s - n_t$
by comparing \FptBCJ\ with \quatrobcj. That is, $n_s = -K_{21}, n_t = K_{23}$
and $n_u =  - K_{21} - K_{23}$.

Therefore, computing the integrals appearing in the five-point scattering amplitude
for each of the twelve regions of integration should provide the remaining five-point
BCJ identities in
a supersymmetric fashion. For example, the partial amplitude $A_{\rm YM}(1,4,2,3,5)$
is obtained by integrating \expre\ over $1\le z_2 \le z_3 \le \infty$, and in this case
the kinematic factors for the different poles appearing in the last equation of
\fiveptbcj\ will be given by combinations of the factors already present in \AYMd, so
that new identities will have to arise. 
In fact, using the transformations $y_3 = (z_3 -1)/z_3$ and $y_2 = (z_2 -1)/z_2$ the integrals
$$
\int_1^{\infty} dz_3 \int_1^{z_3} dz_2 z_3^{\a_{13}}(1-z_3)^{\a_{34}} z_2^{\a_{12}}(1-z_2)^{\a_{24}}
(z_3 - z_2)^{\a_{23}} F(z_3,z_2)
$$
become
$$
\int_0^1 dy_3 \int_0^{y_3} dy_2 y_3^{\a_{34}}(1-y_3)^{\a_{35}} y_2^{\a_{24}}(1-y_2)^{\a_{25}}
(y_3 - y_2)^{\a_{23}} {{\tilde F}(y_3,y_2)\over (1-y_3)^2(1-y_2)^2}
$$
which allow them to be written in terms of $K_j$ and $L_j$ of \MedinaD, provided 
that
\eqn\subs{
\a_{13} \rightarrow \a_{34}, \; \a_{34} \rightarrow \a_{35}, \; \a_{12} \rightarrow \a_{24},\;
\a_{24} \rightarrow \a_{25},\; \a_{51} \rightarrow \a_{14}, \; \a_{23}\rightarrow \a_{23},\;
\a_{45}\rightarrow \a_{51}
}
The only ``new'' integral which is not already computed in \MedinaD\ is the one associated to
$F(z_3,z_2) = (1-z_2)^{-1}(1-z_3)^{-1}$, namely
\eqn\new{
\int_0^1 dy_3 \int_0^{y_3} dy_2 y_3^{\a_{34}}(1-y_3)^{\a_{35}} y_2^{\a_{24}}(1-y_2)^{\a_{25}}
(y_3 - y_2)^{\a_{23}} {1\over y_2y_3(1-y_3)(1-y_2)}.
}
However, \new\ is easily seen to be equal to $K'_1 + K_3 + L_3 \equiv L_8$. Finally, the amplitude
\expre\ integrated over $1\le z_2 \le z_3 \le \infty$ is given by
\eqn\massa{
A(1,4,2,3,5) = L_{2131}{\tilde K}'_1 + L_{2134}{\tilde L}'_1 - L_{2434}{\tilde L}_8
- L_{2431}{\tilde L}'_3 - L_{2331}{\tilde K}'_5 + L_{2334}{\tilde L}_7 + L_{23}(1+\a_{23}){\tilde K}_6
}
where the tildes mean that the substitution \subs\ must be performed. Using the explicit results of
\MedinaD\ for the integrals, the field theory limit of \massa\ is given by
$$
A_{\rm YM}(1,4,2,3,5) = {(L_{2131}+L_{2134}-L_{2434}-L_{2431})\over \a_{14}\a_{35}} - {L_{2434}\over \a_{24}\a_{51}}
+ {(L_{2334} - L_{2331})\over \a_{14}\a_{23}}
$$
\eqn\cool{
- {(L_{2434} + L_{2431})\over \a_{24}\a_{35}}
+ {L_{2334} \over \a_{23}\a_{51}} + L_{23}\({1\over \a_{14}} + {\a_{35}\over \a_{14}\a_{23}} - {\a_{34}\over \a_{23}\a_{51}}\).
}
With the redefinitions of \redef, the contact terms are completely absorbed and \cool\ becomes
$A_{\rm YM}(1,4,2,3,5) =$
$$
 = {({\tilde L}_{2131} + {\tilde L}_{2134}-L_{2434}-L_{2431})\over \a_{14}\a_{35}} - {L_{2434}\over \a_{24}\a_{51}}
+ {({\tilde L}_{2334} - {\tilde L}_{2331})\over \a_{14}\a_{23}}
- {(L_{2434} + L_{2431})\over \a_{24}\a_{35}}
+ {{\tilde L}_{2334} \over \a_{23}\a_{51}}.
$$
From exchanging $2\leftrightarrow 3$ in \cool\ and using \redef\ it follows that
$A_{\rm YM}(1,4,3,2,5) =$
\eqn\neweq{
 = {( L_{3121} +  L_{3124}-{\tilde L}_{3424} + {\tilde L}_{2134})\over \a_{14}\a_{25}} - {{\tilde L}_{3424}\over \a_{34}\a_{51}}
- {({\tilde L}_{2334} - {\tilde L}_{2331})\over \a_{14}\a_{23}}
- {({\tilde L}_{3424} - {\tilde L}_{2134})\over \a_{34}\a_{25}}
- {{\tilde L}_{2334} \over \a_{23}\a_{51}}
}
Finally, comparing the above with \fiveptbcj\ results in the new identifications
$$
n_7 = {\tilde L}_{2331} - {\tilde L}_{2334}, \quad n_{13} = L_{2434} + L_{2431}, \quad n_{14} =
{\tilde L}_{2131} + {\tilde L}_{2134} - L_{2434} - L_{2431},
$$
$$
n_6 = L_{3121} +  L_{3124} - {\tilde L}_{3424} + {\tilde L}_{2134}, \quad n_8 = - L_{3424} + {\tilde L}_{2134}
$$
and therefore the following BCJ identities are obtained
$$
n_6 = n_{15} + n_{10} + n_5 - n_3, \quad n_8 = - n_3 + n_5
$$
\eqn\remain{
n_7 = - n_4 + n_2, \quad n_{13} = n_{11} - n_{10}, \quad n_{14} = n_1 - n_3 - n_{11} + n_{10}.
}
Solving \remain\ and \psbcj\ in terms of $n_1, {\ldots} ,n_6$ gives
$$
n_7 = n_2 - n_4, \quad n_8 = -n_3 + n_5, \quad n_{10} = -n_1+n_3+n_4-n_5+n_6,\quad
n_{11} = n_2 - n_5,
$$
$$
 n_{13} = n_1 + n_2 -n_3 -n_4 - n_6, \quad n_{14} = -n_2+n_4+n_6,
\quad n_{14} = n_1 - n_4
$$
and together with $n_8 - n_6 + n_9 = 0$ and $n_{14} + n_{13} - n_{12} = 0$,
which follow as OPE identities using $U^2$ and $U^4$ or $U^3$ and $U^4$ as integrated vertices,
we get the same solution as (4.12) of \bernN.
Therefore the BCJ identities of \bernN\ were obtained from first principles.
And by using the pure spinor formalism and its pure spinor superspace, we have shown that
the BCJ relations are in fact supersymmetric.

\bigskip
{\bf Acknowledgements: }
I deeply thank Pierre Vanhove for reading an early draft and for suggesting the connection
between the identities \depois\ and the BCJ relations, and also for several discussions.
I thank the organizers of the workshop {\it Hidden Structures in Field Theory Amplitudes 2009},
where I presented parts of this work. I also thank John Carrasco and Henrik Johansson for
conversations during the workshop and Stefan Theisen and Nathan Berkovits for reading
the draft.
I acknowledge support by the Deutsch-Israelische
Projektkooperation (DIP H52).


\appendix{A}{Computation of the kinematic factors}
In this section we compute the OPE's appearing in the amplitude \ampli\
to obtain the explicit expression for the kinematic factors $L_{ijkl}$ in pure spinor superspace.

Using the OPE's 
$$
d_\a(z) V(w) \rightarrow  {D_\a V(w)\over z- w}, \quad
\Pi^m(z) V(w) \rightarrow -  {k^m V(w)\over z- w}, \quad
d_\a(z) \Pi^m(w) \rightarrow  {(\g^m\p\t)_\a\over z- w}
$$
$$
d_\a(z)d_\b(w) \rightarrow -  {\g^m_{\a\b}\Pi_m\over z- w}, \quad
\Pi^m(z)\Pi^n(w) \rightarrow -  {\eta^{mn}\over (z - w)^2}, \quad
d_\a(z)\t^\b(w) \rightarrow  {\d^\b_\a\over (z- w)}
$$
$$
d_\a(z)\p\t^\b(w) \rightarrow  {\d^\b_\a\over (z- w)^2},\quad
w(z)_\a\l^\b(w) \rightarrow - {\d^\b_\a\over z - w}, \quad
N^{mn}(z)\l^\a(w) \rightarrow - \half {(\l\g^{mn})^\a\over z- w}
$$
$$
N^{mn}(z)N_{pq}(w) \rightarrow + {4 \over z - w}N^{[m}_{\phantom{m}[p}\d^{n]}_{q]}
- {6 \over (z - w)^2}\d^n_{[p}\d^m_{q]}
$$
and the equations of motion
$$
D_\a A_\b + D_\b A_\a = \g^m_{\a\b} A_m,\quad
D_\a A_m = (\g_m W)_\a + k_m A_\a,
$$
\eqn\SYM{
D_\a{\cal F}_{mn} = 2k_{[m} (\g_{n]} W)_\a, \quad
D_\a W^{\b} = {1\over 4}(\g^{mn})^{\phantom{m}\b}_\a{\cal F}_{mn},\quad
}
a long computation leads to the OPE between two integrated vertices,
$$
U^2(z)U^3(w) \rightarrow {1 \over (z-w)}\Big[ (k^2\cdot A^3)U^2 - (k^3\cdot A^2)U^3
- (W^2\g_mW^3)\Pi^m 
$$
$$
-\p\t^\a D_\a A^2_\b W_3^\b - \Pi^m k^2_m A^2_\a W^\a_3
+\p\t^\a D_\a A^3_\b W_2^\b + \Pi^m k^3_m A^3_\a W^\a_2
$$
$$
+ {1\over 4}(d\g^{mn}W^2){\cal F}^3_{mn} -  {1\over 4}(d\g^{mn}W^3){\cal F}^2_{mn}
- (k^2_m + k^3_m)(W^2\g_n W^3)N^{mn} + {\cal F}^2_{ma}{\cal F}^3_{na}N^{mn}\Big]
$$
\eqn\OPEint{
+ {1 \over (z-w)^2}\(1 + (k^2\cdot k^3)\)\big[ (A^2W^3) + (A^3W^2) - (A^2\cdot A^3)\big].
}
where we dropped the total derivative terms with respect to $z_2$ which appear when Taylor expanding
the superfields in the double pole. The super-Yang-Mills equations
of motion \SYM\ have been used judiciously to arrive at the simple answer \OPEint.
For example, the terms which contribute to the double pole are given by,
$$
 - A^2_\a [\p\t^\a(z)d_\b(w)] W_3^\b - W^\a_2 [d_\a(z) \p\t^\b(w)] A^3_\b + [\Pi^m(z)\Pi^n(w)] A^2_m A^3_n
$$
$$
 + [d_\b(w)A^2_m(z)][\Pi^m(z) W_3^\b(w)] 
 - [\Pi^n(w)W_2^\a(z)][ d_\a(z) A^3_n(w)]
 + [\Pi^m(z)A^3_n(w)] [\Pi^n(w) A^2_m(z)] 
$$
$$
- [d_\b(w) W_2^\a(z)][ d_\a(z) W_3^\b(w)] + {1\over 4}[N^{mn}(z)N^{pq}(w)]{\cal F}^2_{mn}{\cal F}^3_{pq}.
$$
Using the OPE's one obtains (omitting $(z-w)^{-2}$)
$$
(A^2 W^3) + (A^3 W^2) - (A^2\cdot A^3) + k^3_m (D_\b A^2_m) W_3^\b + k^2_m (D_\a A^3_m) W_2^\a
$$
\eqn\ugly{
- (k^3\cdot A^2)(k^2\cdot A^3) 
+ {1\over 16}\tr(\g^{mn}\g^{pq}){\cal F}^2_{mn}{\cal F}^3_{pq} + {3\over 2}{\cal F}^2_{mn}{\cal F}_3^{mn},
}
where the last term comes from the level $-3$ double pole of the pure spinor Lorentz currents.
One can now use $D_\a A_m = (\g_m W)_\a + k_m A_\a$ and the fact that $k_m (\g^m W)_\a = 0$ to 
simplify \ugly\ to,
$$
(1+ (k^2\cdot k^3))\big[(A^2 W^3) + (A^3 W^2)\big] - (A^2\cdot A^3) - (k^3\cdot A^2)(k^2\cdot A^3) 
- \half {\cal F}^2_{mn}{\cal F}_3^{mn}
$$
\eqn\double{
= (1+ (k^2\cdot k^3))\big[(A^2 W^3) + (A^3 W^2) - (A^2\cdot A^3)\big]
}
where we used $\tr(\g^{mn} \g_{pq}) = - 32 \d^{mn}_{pq}$ and 
$-\half (F^2\cdot F^3) = - (k^2\cdot k^3)(A^2\cdot A^3) + (k^2\cdot A^3)(k^3\cdot A^2)$.

Using the same kind of manipulations as \MafraAR\ one can also prove the following OPE
identity as $z_2 \rightarrow z_1$
\eqn\theorem{
\langle (\l A^1)(z_1)U^2(z_2) {\cal M}\rangle =
-{1\over z_{21}}\langle \big[ A^1_m (\l\g^m W^2) + (\l A^1)(k^1\cdot A^2)\big] (z_1){\cal M} 
+ (A^1 W^2)(z_1)Q{\cal M} \rangle
}
where ${\cal M}(x,\t)$ is any superfield. Furthermore, if $Q{\cal M}=0$ then the following holds true
$$
\langle (\l A^1)(z_1)\big[{1\over 4}(d\g^{mn}W^2){\cal F}^3_{mn}
- {1\over 4}(d\g^{mn}W^3){\cal F}^2_{mn} + {\cal F}^2_{ma}{\cal F}^3_{na}N^{mn}\big](z_3){\cal M}\rangle
$$
\eqn\pu{
= + \langle {1\over 4}(\l\g^p \g^{mn}W^3)A^1_p {\cal F}^2_{mn}{\cal M}
+ {1\over 2} k^2_m( A^1\g^{mn} W^3)(\l \g_n W^2){\cal M}\rangle - (2 \leftrightarrow 3)
}
Also,
\eqn\pd{
-\langle (\l A^1)(k^2_m + k^3_m)(W^2\g_n W^3) N^{mn} M\rangle =
{1\over 2(z_3-z_1)}\langle (\l\g^{mn}A^1)(k^2_m + k^3_m)(W^2\g_n W^3)M\rangle
}
and
$$
\langle (\l A^1)(z_1)(k^2\cdot A^3)U^2(z_3)\rangle =
-{1\over z_3 - z_1}\langle (A^1_m (\l\g^m W^2)+(\l A^1)(k^1\cdot A^2))(k^2\cdot A^3)\rangle
$$
\eqn\pt{
-{1\over z_3 - z_1}\langle (A^1 W^2)((k^2\cdot k^3)(\l A^3)+k^2_m(\l\g^m W^3))\rangle.
}
One can also show by using gamma matrix identities, the pure spinor constraint 
and the SYM equations of motion \SYM\ that
$$
+{1\over 2}k^2_m(A^1\g^{mn} W^3)(\l\g^n W^2)
-{1\over 2}k^3_m(A^1\g^{mn} W^2)(\l\g^n W^3)
+{1\over 2}(\l\g^{mn}A^1)(k^2_m + k^3_m)(W^2\g_n W^3)
$$
\eqn\dif{
+ (A^1W^3)k^3_m(\l\g^m W^2)
- (A^1W^2)k^2_m(\l\g^m W^3) = 0.
}
From \pu, \pd, \pt, \dif\ and the expression for the double pole \double\ we finally get
$$
\langle (\l A^1)(\l A^4)(\l A^5) [U^2U^3](z_3)\rangle =
$$
$$
+ {1\over z_{23}z_{31}}\Big[\langle \big[{1\over 4}(\l\g^p \g^{mn} W^3)A^1_p{\cal F}^2_{mn}
- {1\over 4}(\l\g^p \g^{mn} W^2)A^1_p{\cal F}^3_{mn}\big](\l A^4)(\l A^5)\rangle
$$
$$
+ \langle (A^1_m(\l\g^m W^2) + (\l A^1)(k^1\cdot A^2))(k^2\cdot A^3)(\l A^4)(\l A^5)\rangle
$$
$$
- \langle (A^1_m(\l\g^m W^3) + (\l A^1)(k^1\cdot A^3))(k^3\cdot A^2)(\l A^4)(\l A^5)\rangle
$$
$$
+ \langle \big[k^1_m (\l A^1)(W^2\g^m W^3) + (k^2\cdot k^3)(A^1 W^3)
- (k^2\cdot k^3)(A^1 W^2)\big] (\l A^4)(\l A^5)\rangle
$$
$$
+ {1 \over  z_{23}^2}\langle 
(\l A^1)(\l A^4)(\l A^5)\big[ (A^2W^3) + (A^3W^2) - (A^2\cdot A^3)\big]
(1 + (k^2\cdot k^3))\rangle
- (1\leftrightarrow 4)
$$
from which the following expressions can be read for  $L_{2331}$ and $L_{2314}$,
$$
L_{2331} = A^1_m{\cal F}^2_{mn}(\l\g^n W^3)(\l A^4)(\l A^5)
- {1\over 2}(\l\g_m W^1)(W^2\g^m W^3)(\l A^4)(\l A^5)
$$
$$
       + \big[A^1_m(\l\g^m W^3) + (\l A^1)(k^1\cdot A^3)\big](k^3\cdot A^2)(\l A^4)(\l A^5)\rangle
$$
\eqn\Ldttu{
       + (k^2\cdot k^3) (A^1 W^3)(\l A^2) (\l A^4)(\l A^5)\rangle
       + (k^1\cdot k^2) (A^2 W^3)(\l A^1) (\l A^4)(\l A^5)\rangle - (2\leftrightarrow 3)
}
and 
$$
L_{2314} = (1+ (k^2\cdot k^3))
\langle (\l A^1)(\l A^4)(\l A^5)\big[(A^2W^3) + (A^3W^2) - (A^2\cdot A^3)\big]\rangle
$$
where we used that
$$
 \big[{1\over 4}(\l\g^{p}\g^{mn}W^3)A^1_p{\cal F}^2_{mn} - {1\over 4}(\l\g^{p}\g^{mn}W^2)A^1_p{\cal F}^3_{mn}
  + k^1_m(\l A^1)(W^2\g^{m}W^3) \big](\l A^4)(\l A^5)\rangle 
$$
\eqn\Mariposa{
  = A^1_m{\cal F}^2_{mn}(\l\g^n W^3)(\l A^4)(\l A^5) - {1\over 2}(\l\g_m W^1)(W^2\g^m W^3)(\l A^4)(\l A^5)
  - (2\leftrightarrow 3),
}
which can be checked by writing $k^1_m (\l A^1) = QA^1_m - (\l\g^m W^1)$ in the last term of the LHS and
integrating the BRST charge by parts. 

The expression for $L_{2131}$ can be deduced from
the OPE as $z_2\rightarrow z_1$ followed by $z_3\rightarrow z_1$. Using \theorem\ we
obtain the singularity as $z_2\rightarrow z_1$ 
$$
-{1\over z_{21}}\langle \big[A^1_m(\l\g^m W^2) + (k^1\cdot A^2)(\l A^1)\big](z_1)U^3(z_3)(\l A^4)(\l A^5)\rangle
$$
$$
- {1\over z_{21}}\langle (A^1 W^2)(z_1)\p(\l A^3)(z_3)(\l A^4)(\l A^5)\rangle
$$
whose OPE computation for $z_3\rightarrow z_1$ implies, after some manipulations in superspace, that
$$
L_{2131} = \big[ A^1_m(\l\g^m W^2) + (\l A^1)(k^1\cdot A^2)\big](\l A^4)(\l A^5)((k^1+k^2)\cdot A^3)
$$
$$
- (W^1\g^m W^2)(\l\g_m W^3)(\l A^4)(\l A^5)
+ (A^1\cdot A^2)k^2_m(\l\g^m W^3)(\l A^4)(\l A^5)
$$
$$
+ A^1_m(\l\g^m W^3)(k^1\cdot A^2)(\l A^4)(\l A^5)
- A^2_m(\l\g^m W^3)(k^2\cdot A^1)(\l A^4)(\l A^5)
$$
$$
+ (k^1\cdot k^2)(A^1W^3)(\l A^2)(\l A^4)(\l A^5)
- (k^1\cdot k^2)(A^2W^3)(\l A^1)(\l A^4)(\l A^5)
$$
\eqn\Ldutu{
+ (k^1\cdot k^3)(A^1W^2)(\l A^3)(\l A^4)(\l A^5)
+ (k^2\cdot k^3)(A^1W^2)(\l A^3)(\l A^4)(\l A^5),
}
while $L_{2434}$ and $L_{3121}$ are obtained by exchanging $1\leftrightarrow 4$ 
and $2\leftrightarrow 3$, respectively.

The kinematic factor $L_{2134}$ is given by the coefficient of the OPE
$$
\langle (\l A^1)(z_1)(\l A^4)(z_4)(\l A^5)(z_5) U^2(z_2) U^3(z_3) \rangle
$$
as $z_2 \rightarrow z_1$ followed by $z_3\rightarrow z_4$.
Using \theorem\ the first limit becomes
$$
- {1\over z_{21}}\langle \big[ A^1_m(\l\g^m W^2) + (\l A^1)(k^1\cdot A^2)\big](z_1)(\l A^4)(z_4)(\l A^5)(z_5) U^3(z_3)\rangle
$$
$$
- {1\over z_{21}}\langle (A^1 W^2)(z_1)(\l A^4)(z_4)(\l A^5)(z_5) \p(\l A^3)(z_3)\rangle.
$$
and using \theorem\ again to evaluate as $z_3\rightarrow z_4$ we obtain
$$
+ {1\over z_{21}z_{34}}\langle \big[ A^1_m(\l\g^m W^2) + (\l A^1)(k^1\cdot A^2)\big]
\big[ A^4_m(\l\g^m W^3) + (\l A^4)(k^4\cdot A^3)\big] (\l A^5)\rangle
$$
\eqn\inter{
+ {1\over z_{21}z_{34}}\Big[(k^1\cdot k^2)\langle (A^4 W^3)(\l A^1)(\l A^2) (\l A^5)\rangle
+ (k^3\cdot k^4)\langle (A^1 W^2)(\l A^3)(\l A^4) (\l A^5)\rangle\Big],
}
where we used $QU^3 = \p (\l A^3) = (\p\l^\a) A^3_\a + \Pi^m k^3_m(\l A^3) + \p\t^\a D_\a (\l A^3)$ and
that $Q\big[ A^1_m(\l\g^m W^2) + (\l A^1)(k^1\cdot A^2)\big] = - (k^1\cdot k^2)(\l A^1)(\l A^2)$.
From \inter\ we get the expression for $L_{2134}$,
$$
L_{2134} = 
\langle \big[A^1_m(\l\g^m W^2) + (\l A^1)(k^1\cdot A^2)\big]\big[A^4_m(\l\g^m W^3) 
+ (\l A^4)(k^4\cdot A^3)\big](\l A^5)\rangle
$$
$$
+ (k^1\cdot k^2)\langle (A^4 W^3)(\l A^1)(\l A^2) (\l A^5)\rangle
+ (k^3\cdot k^4)\langle (A^1 W^2)(\l A^3)(\l A^4) (\l A^5)\rangle
$$

\subsec{Explicit proof of $L_{2331} = L_{3121} - L_{2131}$}

From the expressions \Ldttu\ and \Ldutu\ ($L_{3121}$ is obtained from $L_{2131}$ by
exchanging $(2\leftrightarrow 3)$) one can immediately check the following
pure spinor superspace identity
\eqn\import{
L_{2331} = L_{3121} - L_{2131}.
}
To see this first note that all terms containing $(k^i\cdot k^j)$ trivially match on
both sides of \import. Using that $(\l\g^m W^2)(W^3\g_m W^1) + (\l\g^m W^3)(W^1\g_m W^2) 
= (\l\g^m W^1)(W^2\g_m W^3)$ we get, after some trivial cancellations,
$$
 L_{3121} - L_{2131} - L_{2331} =
 -  (\l A^1)(\l A^4)(\l A^5)(k^1\cdot A^2)(k^1\cdot A^3)
$$
$$
+ \big[ A^2_m(\l\g^m W^3)(k^2\cdot A^1)  - (A^1\cdot A^2)k^2_m(\l\g^m W^3)
- A^1_m (\l\g^n W^3){\cal F}^2_{mn} \big](\l A^4)(\l A^5) 
- (2\leftrightarrow 3)
$$
which after using ${\cal F}^2_{mn} = k^2_m A^2_n - k^2_n A^2_m$ is equal to zero,
as we wanted to show.

\appendix{B}{A different pure spinor superspace expression for $A_{F^4}$}

A different superfield expression for \AFq\ may be suggested using the following
argument. The {\it one-loop} amplitude of five massless states
must factorize correctly in the massless poles, which appear when the
surface degenerates into a one-loop four-point amplitude connected
to a three-point at tree-level.  This same factorization
of the five-point one-loop amplitude probes the non-linear expansion
(with five fields) of the one-loop interaction $F^4$.
But the kinematic factors of four-point amplitudes at 
one-loop and tree-level are proportional, therefore the result of this
factorization should also be captured by the tree-level massless five-point amplitude
at the correct $\a'$ order. This is given by the $A_{F^4}$ superfield
above. As discussed in \nonMin, the factorization in the $(12)$-channel ($(23)$-channel)
is given by $L_{12}/\a_{12}$ ($K_{23}/\a_{23}$), where
$$
L_{12} = 
       - 40 \big[ A^1_p (\l\g^p W^2) + (\l A^1) (k^1\cdot A^2)\big](\l\g^m W^5)(\l\g^n W^3){\cal F}^4_{mn}
$$
$$
       + 20 (k^1\cdot k^2)(A^1\g_{mn} W^4)(\l A^2)(\l\g^m W^5)(\l\g^n W^3)
$$
$$
K_{23} = 
       - 40\big[(\l\g^m W^2)(k^2\cdot A^3) -{1\over 4}(\l\g^m\g^{rv} W^3){\cal F}^2_{rv} \big] 
       (\l\g^n W^1)(\l A^4){\cal F}^5_{mn}
$$
$$
       + 20(k^2\cdot k^3)(A^4\g_{mn}W^5)(\l A^3)(\l\g^m W^2)(\l\g^n W^1) 
- (2\leftrightarrow 3)
$$
Therefore it could be argued that $A_{F^4}$ should be proportional to 
the linear combination 
$L_{12}/\a_{12} + K_{23}/\a_{23} + K_{34}/\a_{34} + K_{45}/\a_{45}
+ L_{51}/\a_{51}$. One can check that the {\it bosonic} components satisfy
\eqn\AFqd{
A_{F^4}(\t) = - {1\over 40}\( {L_{12} \over \a_{12}} + {K_{23}\over \a_{23}} 
+ {K_{34}\over \a_{34}} + {K_{45}\over \a_{45}}
+ {L_{51}\over \a_{51}} \).
}

\appendix{C}{The MHV amplitude}

It is interesting to (formally) rewrite our component expansions in the
language of four-dimensional helicity formalism. If the helicities
of the gluons are $(--+++)$ we use the following
conventions,
$$
e^I_{\a{\dot \a}} = \sqrt{2}{\psi^I_\a {\bar \chi}^I_{\dot \a}\over [{\bar \psi}^I{\bar \chi}^I]}, \quad I=1,2, \quad
e^J_{{\dot \b}\b} = \sqrt{2}{{\bar \psi}^J_{\dot \b}  \chi^J_\b \over \langle  \chi^J \psi^J \rangle}, \quad J=3,4,5
$$
where $\langle \psi \chi\rangle = \psi^\a \chi_\a = \e^{\a\b}\psi_\b\chi_\a$ and
$[{\bar \psi}{\bar \chi}] = {\bar \psi}_{\dot \a}{\bar \chi}^{\dot \a} 
= \e_{{\dot \a}{\dot \b}}{\bar \psi}_{\dot \b}{\bar \chi}^{\dot \a}$
are the spinor products and $\ang{ij}[ij] = -2 \a_{ij}$.
For the specific choice of reference momenta $(2,1,1,1,1)$ they imply
$$
 (e^1\cdot e^3) =
 (e^1\cdot e^4) =
 (e^1\cdot e^5) = 
 (e^3\cdot e^4) =
 (e^3\cdot e^5) = 
 (e^4\cdot e^5) = 0
$$
$$
 (k^2\cdot e^1) = 
        (k^1\cdot e^2) =
        (k^1\cdot e^3) = 
        (k^1\cdot e^4) =
        (k^1\cdot e^5) = 0
$$
and one can check \PSS\ that
$L_{12} = L_{51} = L_{2131} = L_{2134} = L_{2314} = 0$.
With this gauge choice the superfields \AYM\ and \AFq\ become
$$
A_{\rm YM}(\t) =
- {L_{3424}\over \a_{34}\a_{51}}
- {L_{2334}\over \a_{23}\a_{51}} 
- {L_{2331}\over \a_{23}\a_{45}}
$$
$$
A_{F^4}(\t) =
 -{1\over 40}\big[ 
 {K_{23}\over \a_{23}} + {K_{34}\over \a_{34}} + {K_{45}\over \a_{45}}
\big]
$$
where
$$
(L_{2331}, L_{3424}, L_{2334})  = - {\sqrt{2} \ang{12}^4 \over 23040 \a_{12}}\( 
{  [23]^2 [45]\over \ang{14}\ang{15}},
{  [25]^2 [34]\over \ang{13}\ang{14}},
-{ [23][24] [45]\over \ang{13}\ang{14}}
\)
$$
$$
K_{23} = + {\sqrt{2} \over 576}{ \ang{12}^3 [23] [45]^2\over \ang{13} }
= - {\sqrt{2}\over 72}\a_{23}\a_{45}\ang{12}^4\big[ {\a_{51}\over N(12345)} + {\a_{25}\over N(12543)}\big]
$$
$$
K_{34} = + {\sqrt{2} \over 576}{ \ang{12}^4 [25]^2 [34]\over \ang{13}\ang{14}} 
= - {\sqrt{2}\over 72}\a_{25}\a_{34}\ang{12}^4\big[ {\a_{35}\over N(12534)} - {\a_{45}\over N(12543)}\big]
$$
$$
K_{45} = + {\sqrt{2} \over 576}{ \ang{12}^4 [23]^2 [45]\over \ang{14}\ang{15}} 
= - {\sqrt{2}\over 72}\a_{23}\a_{45}\ang{12}^4\big[ {\a_{34}\over N(12345)} - {\a_{35}\over N(12354)}\big],
$$
where $N(ijklm) = \ang{ij}\ang{jk}\ang{kl}\ang{lm}\ang{mi}$.
Using the results above it is straightforward to obtain, in the NS sector,
$$
A_{\rm YM} = {\cal M}_{\rm MHV} = {\sqrt{2}\over 2880}{\ang{12}^4 \over \ang{12}\ang{23}\ang{34}\ang{45}\ang{51}},
$$
which agrees with the well-known MHV amplitude up to an overall coefficient.
The superfield expression for $A_{F^4}$ becomes
$$
A_{F^4} =  {\cal M}_{\rm MHV}\big[ \a_{45}\a_{51} + \a_{23}\a_{34}
+ \a_{25}\a_{35}{N(12345)\over N(12534)} - \a_{23}\a_{35}{N(12345)\over N(12354)}\big],
$$
which can be rewritten as 
\eqn\minus{
A_{F^4} =  {\cal M}_{\rm MHV}\big[ \a_{45}\a_{51} + \a_{23}\a_{34}
- [12]\ang{23}[35]\ang{51}
\big],
}
agreeing with (5.45) of \vonkMHV\ and (37) of \stieMulti, 
apart from the overall coefficient. 

\appendix{D}{The integrals $K_j$}

In \MedinaD\MedinaC\ the following identities  were derived\foot{We
use a different convention where $\p_m = k_m$ instead of
$\p_m = ik_m$. Therefore one must replace 
$\a_{ij} \rightarrow  -\a_{ij}$ in the identities of \MedinaD. The only place where it matters
is the identity involving $K_6$.}
$$
\a_{34}K_2 = \a_{13}K_1 + \a_{23}K_4, \quad \a_{24}K_3 = \a_{12}K_1 - \a_{23}K_5, \quad K_1 = K_4 - K_5
$$
$$
\a_{12}K_2 = \a_{24}K'_1 + \a_{23}K'_4, \quad \a_{13}K_3 = \a_{34}K'_1 - \a_{23}K'_5,  \quad K'_1 = K'_4 - K'_5
$$
\eqn\relat{
(1+\a_{23})K_6 = \a_{34}K'_4 - \a_{13}K_5
= \a_{12}K_4 - \a_{24}K'_5,
}
and their explicit expansions in terms of $\a_{ij}$ were computed at length.
However, as mentioned in \MedinaC, by defining
\eqn\T{
T = \a_{12}\a_{34}K_2 + (\a_{12}\a_{51} - \a_{12}\a_{34} + \a_{34}\a_{45})K_3,
}
all integrals $K_j$ and $K'_j$ can be written in terms of $T$ and $K_3$.
For example, from \relat\ one can check that (and similarly for $K'_j$)
$$
K_1 = {\a_{34}\over \a_{45}}K_2 + {\a_{24}\over \a_{45}}K_3,
$$
$$
K_4 = {\a_{12}\a_{34}K_2 + \a_{23}\a_{34}K_2 - \a_{13}\a_{24}K_3\over \a_{23}\a_{45}}, \hskip 0.2cm
K_5 = {\a_{12}\a_{34}K_2 - \a_{13}\a_{24}K_3 - \a_{23}\a_{24}K_3\over \a_{23}\a_{45}}
$$
which imply
\eqn\Ku{
K_1 = {T\over \a_{12}\a_{45}} - \({\a_{34}\over \a_{12}} + {\a_{23}\over \a_{45}}\)K_3
}
\eqn\Kq{
K_4 = \({1\over \a_{23}} + {1\over \a_{12}}\){T\over \a_{45}} - \(
{\a_{51}\over \a_{23}} + {\a_{34}\over \a_{12}} - {\a_{13}\over \a_{45}}\)K_3
}
\eqn\Kc{
K_5 = {T\over \a_{23}\a_{45}} - \({\a_{12}\over \a_{45}} + {\a_{51}\over \a_{23}} -1 \)K_3
}
\eqn\mila{
(\a_{23} + 1)K_6 =
 (1 - {\a_{24}\over \a_{51}} - {\a_{13}\over \a_{45}}) {T\over \a_{23}}
+ ( \a_{13} \a_{24} - \a_{12} \a_{34} - \a_{23} \a_{34} - \a_{12} \a_{23})
{K_3\over \a_{23}}
}
where we used \MedinaC,
$$
\a_{13} = \a_{45} - \a_{12} - \a_{23}, \quad
\a_{14} = \a_{23} - \a_{51} - \a_{45}, \quad
\a_{24} = \a_{51} - \a_{23} - \a_{34}
$$
\eqn\momconser{
\a_{25} = \a_{34} - \a_{12} - \a_{51}, \quad
\a_{35} = \a_{12} - \a_{45} - \a_{34}.
}
It was shown in \MedinaD\ that 
under the twist $\a_{12} \leftrightarrow \a_{34}$, $\a_{13} \leftrightarrow \a_{24}$, $\a_{23} \leftrightarrow \a_{23}$
the integrals behave as
$$(T,K_1, K_2, K_3, K_4, K_5, K_6) \leftrightarrow (T,K'_1,K_2,K_3,K'_4,K'_5, K_6)$$
which allows one to easily obtain $K'_1$, $K'_4$ and $K'_5$ from \Ku, \Kq\ and \Kc.

\listrefs

\end